\documentclass{article}
\usepackage{spconf,amsmath,graphicx}

\usepackage[hyphens]{url}
\usepackage{subfiles}
\usepackage{color}
\usepackage{booktabs}
\usepackage{amsfonts}
\usepackage{footnote}
\usepackage{hyperref}
\usepackage{graphbox}
\usepackage{float}
\usepackage{multirow}
\usepackage{array}
\usepackage{subcaption}
\usepackage[dvipsnames]{xcolor}
\graphicspath{{images/}{../images/}}

\title{Acoustic-to-Word Recognition with Sequence-to-Sequence Models}
\name{Shruti Palaskar and Florian Metze}
\address{Carnegie Mellon University\\
Language Technologies Institute\\
Pittsburgh, PA, USA\\
\texttt{\{spalaska{\textbar}fmetze\}@cs.cmu.edu}}

\begin{document}
%\ninept
%
\maketitle
\begin{abstract}

Acoustic-to-Word recognition provides a straightforward solution to end-to-end speech recognition without needing external decoding, language model re-scoring or lexicon. While character-based models offer a natural solution to the out-of-vocabulary problem, word models can be simpler to decode and may also be able to directly recognize semantically meaningful units. We present effective methods to train Sequence-to-Sequence models for direct word-level recognition (and character-level recognition) and show an absolute improvement of 4.4-5.0\% in Word Error Rate on the Switchboard corpus compared to prior work. In addition to these promising results, word-based models are more interpretable than character models, which have to be composed into words using a separate decoding step. We analyze the encoder hidden states and the attention behavior, and show that location-aware attention naturally represents words as a single speech-word-vector, despite spanning multiple frames in the input. We finally show that the Acoustic-to-Word model also learns to segment speech into words with a mean standard deviation of 3 frames as compared with human annotated forced-alignments for the Switchboard corpus.

\end{abstract}
\begin{keywords}
end-to-end speech recognition, encoder-decoder, acoustic-to-word, speech embeddings
\end{keywords}
\section{Introduction}

\label{sec:intro}
%This also enables us to include lengthier utterances as there is a less memory overhead in the attention computation. 

Direct Acoustic-to-Word (A2W) mapping is relevant for Automatic Speech Recognition (ASR) because it no longer requires a lexicon or separately-trained language model, that is currently necessary to decode in a grapheme, phoneme, character or sub-word models. It would lead to truly end-to-end speech recognition models giving P(Words$|$Acoustics). Another strong motivation for building direct word models is to obtain semantically meaningful representations that would be useful in co-learning tasks with other modalities like text where the unit of representation is generally words. 

Recently, A2W models have been explored in order to have a simpler and efficient solution for end-to-end recognition. 
The challenges of these models are how to handle the large word vocabularies without requiring thousands of hours of data. As a way to manage that problem, \cite{ibm_building_competitve} restricts the vocabulary to 10,000 frequently occurring words or resorts to using sub-word units to avoid out-of-vocabulary (OOV) words. But both of these approaches have certain drawbacks. First, restricted vocabulary leads to OOV words. And in addition to that, sub-word units are not semantically or syntactically as rich as whole-word units. In this paper, we present a large vocabulary full word model that does not have these constraints.

% Todo: why "until recently"? Did something change fundamentally? I.e. a new dataset came out? I think that simply people are investigating A2W models because they can be even simpler, faster, etc - and they (i.e. we) are trying to close the performance gap.

For end-to-end speech recognition, Recurrent Neural Network (RNN) acoustic models paired with Connectionist Temporal Classification (CTC) \cite{ctc,graves2014asr} or attention-based Sequence-to-Sequence (S2S) models \cite{seq2seq,chorowski2014,attn_is_all,asr_bpe_google} provide a more powerful and simple framework that scales well with large training corpora. These sequence-based models no longer need predefined alignments between acoustic features and transcriptions, and this flexibility makes working towards A2W models a possibility now.

Recently, notable progress has been made towards building direct A2W models using CTC \cite{ibm-2017-swbd,microsoft-2018-a2w,hasim-2016-a2w,zenkel_subword} but it either requires large training data \cite{microsoft-2018-a2w,hasim-2016-a2w, google_asr_sota} or smaller vocabulary \cite{ibm_building_competitve,ibm-2017-swbd,zenkel_subword}. In this paper, we present one such approach using no more than 300 hours of training data but with a S2S model instead of a CTC model. 

Our motivation for using a S2S model is that while CTC follows a monotonic alignment between acoustic frames and word predictions, an A2W model would benefit more from a flexible alignment scheme between acoustics and words. As spoken words are more variable in a number of frames per word as compared to the number of frames for characters or phonemes, enforcing a monotonicity for such alignment would be an overhead. Especially in spontaneous speech, or noisy, or multi-speaker scenarios, the labels may not be 100\% accurate and model will benefit with flexibility. With this method, we can use direct acoustic-to-word modeling in other sequence-based tasks like part-of-speech tagging or syntactic parsing with speech input.  

An A2W model may show better interpretability as it directly maps two correlated streams of data, end-to-end i.e. speech to words. In this paper, we explore the interpretability, in particular, we analyze the encoder hidden states and the attention mechanism of a S2S model. In the process, we find that this model learns to segment input speech into words, silence, and non-silence parts without any added supervision for this segmentation. We also find that the word-alignments produced by our model are as accurate as human annotated segmentations. Using this learned segmentation, we are able to directly obtain speech-word-vectors from these models.

\section{Related Work}

A2W modeling has been largely pursued using CTC models. Soltau et al. \cite{hasim-2016-a2w} introduced the first A2W model but needed a very large training corpus (125,000 hours) due to the large vocabulary of A2W models, 100000 words in their case. These word vocabularies are noticeably much larger than character or sub-word vocabularies. Audhkhasi et al. \cite{ibm-2017-swbd} found that filtering out rare words and replacing them with an OOV symbol alleviates the need for large data. But producing OOV would lead to higher word error rates. A common technique to solve this OOV problem has been to use word-level prediction for frequent words and revert to the character or sub-word prediction for rare words \cite{microsoft-2018-a2w,ibm_building_competitve}. This is a two-step procedure, straying from the regular sequence-to-sequence mapping of acoustics to units. Recently, Li et al. \cite{microsoft-2018-a2w} proposed a hybrid-CTC model where an A2W model consulted an Acoustic-to-Letter model upon generation of the OOV symbol. They also proposed a mixed-unit CTC model using frequent words, letters and sub-words although again with large amounts of data (approximately 3,400 hours). In \cite{ibm_building_competitve}, the authors propose a Spell and Recognize model where they first predict the ``spelling" of the word before composing it into a word unit (if a frequent word), or preserving the ``spelling" as character units. This approach is single-step method as they use a common softmax for the mixed vocabularies. All these methods described above use the CTC loss function. 

S2S models have also been used for recognizing sub-word \cite{BPE} or word-piece units in ASR \cite{asr_bpe_chan,google_asr_sota,asr_bpe_google} that no longer have OOV words but these results were presented with 12,500 hours of in-house speech data. Lu et al. \cite{renals_seq2seq_swbd} present one of the first S2S models for large vocabulary A2W recognition with the 300 hour Switchboard corpus with a vocabulary of 30,000 words. In this paper, we build upon their methods and present an effective way of training end-to-end A2W models with improved performance. 

Another area of research that our paper is relevant to is speech-vector representation learning. \cite{speech2vec,kamper_acoustic_word_embeddings,harwath_glass_2015,bengio_speech_embeddings} all explore ways to extract speech embeddings. Their methods are commonly unsupervised learning based on clustering where they do not use the transcripts or do not perform speech recognition. In this work, we obtain similar speech-vectors as a by-product of our speech recognition.

\section{Sequence-to-Sequence Model}

\label{sec:methods}
Our S2S model is similar in structure to the Listen, Attend and Spell model \cite{LAS} which consists of 3 components: the encoder network, a decoder network and an attention model. The encoder maps the input acoustic features vectors $\textbf{x} = (\textbf{x}_1,\textbf{x}_2,...,\textbf{x}_T)$ where $\textbf{x}_i \in \mathcal{R}^d$ into a sequence of higher-level features $\textbf{h} = (\textbf{h}_1,\textbf{h}_2,...,\textbf{h}_{T'})$. The encoder is a multi-layer bi-directional Long Short Term Memory (BLSTM) RNN that is structured as a pyramid by skipping every other frame between certain encoder layers for efficient training. This reduces the length of the input from $T$ to $T'$. This encoder network is analogous to the traditional acoustic model of an ASR. The decoder network is also an LSTM network that learns to model the output distribution over the next target conditioned on sequence of previous predictions i.e. $P(\textbf{y}_l | y_{l-1}^*,y_{l-2}^*,...,y_0^*,\textbf{x})$ where $\textbf{y}* = (y_0^*,y_1^*,...,y_{L+1}^*)$ is the ground-truth label sequence. In this work, $y_i^* \in \mathcal{U}$ can be a token from a character, sub-word or word vocabulary. This decoder network is similar to the language model in traditional ASR as it generates targets $\textbf{y}$ from $\textbf{h}$ using an attention mechanism. The attention model learns an alignment weight vector between the encoding $\textbf{h}$ and the current output of decoder $\textbf{y}_l$. At each time step, the attention module computes a context vector that is fed into the decoder together with the previous ground-truth label $y_{l-1}^*$.

%(we do not use scheduled sampling \cite{scheduled_sampling}). 

%While generating larger target units with an S2S model, longer context within the attention mechanism would also be beneficial. 
We use a location-aware attention mechanism \cite{chorowski2015attention} that enforces monotonicity in the alignments, which may be beneficial for speech recognition. To do so, the location-aware attention applies a convolution across time to the attention of previous time step using trainable filters. This convolved attention feature is used for calculating the attention for the current time step. We apply a one-dimensional convolution $\mathcal{K}$ along the input feature axis $t$ to get a set of $T$ features $\lbrace\textbf{f}\rbrace_{t=1}^T$ described as follows:
\begin{align*}
\lbrace\textbf{f}\rbrace_{t=1}^T &= \mathcal{K} * \textbf{a}_{l-1} \\
e_{lt} &= \textbf{g}^{\textnormal{T}} \textnormal{tanh (Lin}(\textbf{y}_{l-1}) + \textnormal{Lin}(\textbf{h}) + \textnormal{LinB}(\textbf{f}_t)) \\
a_{lt} &= \textnormal{Softmax}(\lbrace e_{lt}\rbrace_{t=1}^{T})
\end{align*}
where $ \textbf{a}_{l-1} = \left[a_{l-1,1},...,a_{l-1,T}\right]^\textnormal{T}$, $\textbf{g}$ is a learnable vector parameter, $\left\lbrace e_{lt}\right\rbrace_{t=1}^T$ is a $T$-dimensional vector, Lin() is a linear layer with learnable matrix parameters without bias vectors, LinB() is a linear layer with learnable matrix and bias parameters.

The S2S model is trained by optimizing the cross entropy loss function which maximizes the log-likelihood of the training data. We use beam search to perform inference. We also apply unigram label smoothing that distributes the probability of most-probable token to prevent the over-confidence of the model \cite{label_smoothing,chorowski_decoding}.

\section{Experimental Setup}

% \subsection{Dataset}
\label{sec:experiments}
We use the standard 300-hour Switchboard corpus (SW, LDC97S62) \cite{switchboard_corpus} which consists of 2,430 two-sided telephonic conversations between 500 different speakers and contains 3 million words of text. We evaluate on the HUB5 eval2000 (LDC2002S09, LDC2002T43) containing Switchboard subset similar to training data and CallHome (CH) subset that is a tougher set. There are 196,656 total utterances out of which we use the first 4,000 utterances as a validation set.
Our input features are 80-dimensional log-mel filter banks normalized with per-speaker mean and variance. We also use 3-dimensional pitch features.

% \subsection{Target Units}
% \label{ssec:target_units}
We present three different types of target units for speech recognition in this paper: characters, BPE units and words. The character vocabulary is made of 46 units containing 26 letters, 10 digits, and other frequently occurring special symbols. We try different BPE vocabularies like 300, 500, 1k, 5k, 10k and 16k. We finally present a large-vocabulary model made of all 29,874 unique words in the Switchboard set. The vocabularies also contain non-language special symbols that denote noise, vocalized-noise and laughter. We train character and word level RNN language models on the Switchboard + Fisher (LDC2004T19) \cite{fisher_corpus} transcripts as is the common practice for this data.

% \subsection{System Description}
% \label{ssec:system_description}
Our encoder consists of 6 layers each with 320 bi-directional LSTM cells. The second and third layer skip every other frame to get a reduction of $T/4$ in input frames. We use the AdaDelta \cite{adadelta} optimizer. The location-aware attention convolution uses 10 filters with width 100. We use a projection layer of 320 dimensions after each layer of the encoder. Our decoder is a single layer LSTM containing 300 cells. We initialize all parameters uniformly within $\left[-0.1,0.1\right]$ unless otherwise specified. We use unigram label smoothing with weight 0.05. The beam size used for all experiments is 10. We use the ESPnet toolkit\cite{suyoun_joint_ctc_attn,shinji_espnet} as a starting point for our experiments.

\section{Results}

In Table \ref{tab:wer_char}, we present our character-level S2S model and compare with previously published CTC and S2S models, using the 300h SW corpus and character vocabularies for better understanding. According to these results, our models obtain the best Word Error Rate (WER) in both SW and CH test sets among the S2S models with and without a language model. We also perform better than all CTC models in the SW test set and the difference in the CH set is minor. Furthermore, we observe a 13\% relative improvement in the SW subset by using an RNNLM with shallow fusion \cite{shallow_fusion} which is trained at the character and word level.

% deep speech contains language model

%%%%%%%%%%%%%%%%%%%%%%%%%%%%%%%%%%%%%%%%%%%%%%%%%%%%%%%%%%%%%%%%%%%%%%%%%%%%
% CHAR SWBD Results
% exp/train_nodup_char_blstmp_e6_subsample1_2_2_1_1_unit320_proj320_d1_unit300_location_aconvc10_aconvf100_mtlalpha0_adadelta_bs48_20180606_005902_lsmunigram0.05
% Char: decode_eval2000_beam10_p0_len0.0-0.0_nj8_rerun
% +LM: decode_eval2000_beam10_p0.1_len0.0-0.0_nj8_rnnlm0.2
% +word LM: decode_eval2000_beam10_p0_len0.0-0.0_nj8_char_word_wordrnnlm0.2
%%%%%%%%%%%%%%%%%%%%%%%%%%%%%%%%%%%%%%%%%%%%%%%%%%%%%%%%%%%%%%%%%%%%%%%%%%%%
\begin{table}[h!]
\caption{Word Error Rate (WER) for the SW and CH test sets using \textbf{character target units}, and comparison with other end-to-end character-level models. We compare with the re-scored character-LM results from prior work when available.}
\centering
\begin{tabular}{ p{3.7cm}  p{1cm} p{0.9cm}  p{0.9cm}  }
\hline
\hline
\hspace*{0.2cm}
			        &           &\multicolumn{2}{c}{\textbf{WER (\%)}}       \\
\textbf{Model}		& \textbf{Vocab}     & \textbf{SW} 	    & 	\textbf{CH} \\ 
			        \hline
			        \hline
%CTC Char 
%\cite{jurafsky_ctc_char}&33     &  38.0     &   56.1                \\
%\hspace{0.3cm} LF-MMI(*) 
%\cite{povey-lfmmi}            &           & 14.4      &   25.2       \\
Prior Work CTC \\
%\hspace{1.2cm}  \\
\hspace{0.3cm} Hannun et al. +LM 
\cite{deep_speech}            & 29        & 20.0      &   31.8       \\
\hspace{0.3cm} Zweig et al. +LM 
\cite{droppo_ctc_char}        & 79        &  19.8     &   32.1       \\
\hspace{0.3cm} Audhkhasi et al. 
\cite{ibm_building_competitve}& 79        & 18.9      &   \textbf{30.9}       \\
Prior Work S2S \\
\hspace{0.3cm} Lu et al. +LM 
\cite{renals_seq2seq_swbd}    & 35        &  32.6     &   51.9       \\
\hspace{0.3cm} Zenkel et al. 
\cite{thomas_decoding}        & 46        & 28.1      &   40.6       \\
\hspace{0.3cm} Toshniwal et al. 
\cite{toshniwal_multitask}    & N/A       &  23.1     &   40.8       \\
                    \hline
                    \hline
Our models \\
\hspace{0.3cm} S2S Char & 46     & \textbf{18.0}&  32.5 \\
\hspace{0.3cm} S2S Char +LM & 46 & \textbf{17.1}&\underline{\textbf{31.1}}\\
\hspace{0.3cm} S2S Char +Word LM &46& \textbf{15.6}&\underline{\textbf{31.0}} \\
%BPE 500             & 572       &  18.0		&   33.3      		 \\
%BPE 1k              & 1069      &  18.8        &   33.5                \\
                    \hline
                    \hline
\end{tabular}
\label{tab:wer_char}
\end{table}
%%%%%%%%%%%%%%%%%%%%%%%%%%%%%%%%%%%%%%%%%%%%%%%%%%%%%%%%%%%%%%%%%%%%%%%%%%%%

%%%%%%%%%%%%%%%%%%%%%%%%%%%%%%%%%%%%%%%%%%%%%%%%%%%%%%%%%%%%%%%%%%%%%%%%%%%%
% WORD SWBD Results
% BPE: exp/train_nodup_bpe12000_blstmp_e6_subsample1_2_2_1_1_unit320_proj320_d1_unit300_location_aconvc10_aconvf100_mtlalpha0_adadelta_bs48_20180702_034555_lsmunigram0.05/decode_eval2000_beam10_p0.1_len0.0-0.0_nj8_rerun

%############### CHANGE THIS we decode word models with beam 5, but others with beam 10 #############
% Wordgt5: exp/train_nodup_word_blstmp_e6_subsample1_2_2_1_1_unit320_proj320_d1_unit300_location_aconvc10_aconvf100_mtlalpha0_adadelta_bs48_20180606_163555_lsmunigram0.05/decode_eval2000_beam5_p0_len0.0-0.0_nj8_rerun

% Wordgt5*: exp/train_nodup_word_blstmp_e6_subsample1_2_2_1_1_unit320_proj320_d1_unit300_location_aconvc10_aconvf100_mtlalpha0_adadelta_bs48_20180607_010358_initchar_lsmunigram0.05/decode_eval2000_beam5_p0_len0.0-0.0_nj8_rerun

% large vocab: exp/train_nodup_wordgt1_blstmp_e6_subsample1_2_2_1_1_unit320_proj320_d1_unit300_location_aconvc10_aconvf100_mtlalpha0_adadelta_bs32_20180705_190034_lsmunigram0.05/decode_eval2000_beam5_p0_len0.0-0.0_nj8_rerun

% large vocab w LM: exp/train_nodup_wordgt1_blstmp_e6_subsample1_2_2_1_1_unit320_proj320_d1_unit300_location_aconvc10_aconvf100_mtlalpha0_adadelta_bs32_20180705_190034_lsmunigram0.05/decode_eval2000_beam5_p0_len0.0-0.0_nj8_rnnlm0.08

%%%%%%%%%%%%%%%%%%%%%%%%%%%%%%%%%%%%%%%%%%%%%%%%%%%%%%%%%%%%%%%%%%%%%%%%%%%%
\begin{table}[h!]
\caption{Word Error Rate (WER) for the SW and CH test sets using BPE and \textbf{word level target units}, and comparison with other end-to-end word-level models. * denotes character initialization} 
\centering
\begin{tabular}{ p{4.4cm}  p{1cm} p{0.7cm}  p{0.7cm}  }
\hline
\hline
\hspace*{0.2cm}
			        &           &\multicolumn{2}{c}{\textbf{WER (\%)}}       \\
\textbf{Model}		& \textbf{Vocab}    & \textbf{SW} 	& 	\textbf{CH} 	 \\ 
			        \hline
			        \hline
Prior Work CTC\\
%\hspace{0.3cm} Zenkel et al. (BPE)
%\cite{zenkel_subword}       &10000  &17.8   & 29.0                \\
\hspace{0.3cm} Audhkhasi et al.
\cite{ibm_building_competitve}&10000 & 14.5  & 23.9                 \\ 
\hspace{0.3cm} Chen et al. 
\cite{chen2018modular}      &29874   & 24.9  & 36.5                 \\
Prior Work S2S\\
\hspace{0.3cm} Chen et al. 
\cite{chen2018modular}      &29874   & 31.2  & 40.5                 \\
\hspace{0.3cm} Lu et al.
\cite{renals_seq2seq_swbd} &29874&  26.8     &   48.2                \\
\hspace{0.3cm} Lu et al. +LM
\cite{renals_seq2seq_swbd} &29874&  26.2     &   47.4                \\
                    \hline
                    \hline
Our models \\
\hspace{0.3cm} S2S BPE 12k & 11690 &  \textbf{21.3}&\textbf{35.7}\\
\hspace{0.3cm} S2S Word $>=$ 5& 11069     &  23.0     &   37.2     \\
\hspace{0.3cm} S2S Word $>=$ 5*& 11069     &  22.4     &   36.1    \\
                     \hline
% \hspace{0.3cm} S2S BPE 16k          & 15239     &  22.9     &   37.1                \\
% \hspace{0.3cm} \textbf{S2S Word $>=$ 3} & 15027 &\textbf{22.8}&\textbf{36.4} \\
%                     \hline
\hspace{0.3cm} S2S Large Vocab & 29874  & \textbf{22.4}& \textbf{36.2} \\
\hspace{0.3cm} S2S Large Vocab + LM & 29874  & \textbf{22.1}& 36.3 \\
                    \hline
                    \hline
\end{tabular}
\label{tab:wer_word}
\end{table}
%%%%%%%%%%%%%%%%%%%%%%%%%%%%%%%%%%%%%%%%%%%%%%%%%%%%%%%%%%%%%%%%%%%%%%%%%%%%

In Table \ref{tab:wer_word}, we present the A2W models with BPE and word units. Our first model consists of words occurring at least 5 times (Word $>=$ 5) in the training set that led to 11069 words but with an OOV rate of 2.3\% in the eval2000 test set. To address this high OOV rate, we tried to match the word vocabulary by an equivalent BPE vocabulary of 12k merge operations. This model performed better than the word model as expected. We also experiment with initializing the word$>=$5 model with a pretrained character model (similar to \cite{ibm_building_competitve}) for better convergence and observe improvements.

Our second model is a large vocabulary model made of all the words in the training set. This model performs better than the previous word model which may be due to absence of the frequently occurring OOV token. We get an absolute improvement of 4.4\% and 12\% in SW and CH subsets over our baseline \cite{renals_seq2seq_swbd} without a language model. Ideally, S2S A2W model does not need a separate language model as it directly predicts a sequence of words using the decoder LSTM. But as the LM is trained on a larger corpus, we integrate it to check its effect and do not observe improvements as large as the character model. %It is interesting to note that the CTC A2W model performs much better than the CTC character-based model, while this is not the case for S2S models yet. 

\textbf{Comparison with CTC.} The vocabularies of CTC models (both character and word) is different than ours hence models are not comparable. Prior work in CTC \cite{ibm_building_competitve} has almost 20,000 less words than our model and they used strong hyper parameter tuning techniques to arrive at a successful A2W model. On a similar setup, their character-based model is worse by 5\% WER. In the paper, they do not provide a reason for this behavior. In our S2S model, we observe the reverse trend i.e. the word-model performs worse than character-model. This is an interesting trend for CTC and S2S models and needs further exploration. We note that CTC and S2S models are not comparable with each other due to critical differences in loss computation.

%We initialize the min count models with a pre-trained character model similar to \cite{ibm_building_competitve} and observe better convergence properties (1\% absolute reduction in WER). OOV rate for the eval2000 test set with min count 5 and 3 models is 2.3\% and 1.9\% respectively. We build BPE models with similar vocabulary sizes as these models to observe difference in WER with and without OOV words. For 12k BPE operations, there

% Language Model integration
% We use a character RNN-LM for the character models and word RNN-LM for the word models. The perplexity of our character LM  is 2.96 while that of our word LM is 54.9 (vocabulary \shruti{fix this}).

\section{Attention Analysis}

In the following two sections we analyze the behavior of S2S models, specifically for the A2W recognition task. We analyze attention in the decoder and the hidden representations of the encoder. % We see that the S2S A2W models learns to segment speech into words and learns good word boundaries. "Don't give out conclusion in the beginning. -Des"

%%%%%%%%%%%%%%%%%%%%%%%%%%%%%%%%%%%%%%%%%%%%%%%%%%%%%%%%%%%%%%%%%%%%%%%%%%%%
% NXT SWBD annotations
% \footnote{\url{https://catalog.ldc.upenn.edu/LDC2009T26}}
%%%%%%%%%%%%%%%%%%%%%%%%%%%%%%%%%%%%%%%%%%%%%%%%%%%%%%%%%%%%%%%%%%%%%%%%%%%%
\textbf{Human Annotated Word Boundaries in SWBD.} NXT Switchboard Annotations (LDC2009T26) are a subset of the Switchboard corpus (LDC97S62) containing 1 million words that were annotated for syntactic structure and disfluencies as part of the Penn Treebank project. This subset of the Switchboard corpus contains human annotated word-level forced alignments that mark the beginning and end of each word in the utterance in time \footnote{\url{http://groups.inf.ed.ac.uk/switchboard/structure.html}}. In the following sections, we analyze attention behavior of the A2W model and the speech-word-vectors obtained from it. To do this analysis, we need groundtruth word-level segmentations and this corpus is a good match.

From NXT Switchboard, we choose those utterances that are also present in the Treebank-3 (LDC99T42) corpus. The speech in this corpus is re-segmented to match the sentences in Treebank-3. We filter out utterances with less than 3 words resulting in 67,654 utterances in total.  This is divided into 56,100 train, 5,829 validation and 5,725 test sets. We train a separate A2W model with this data in the same setup as described in Section \ref{sec:experiments}, without using any explicit information about word-segments. We only train on this dataset to avoid introducing a more variability in our analysis, i.e. are the segmentations due to our model or due to training with a larger corpus (SW 300h)? In our setup, we split compound words into two words (eg. they`re $\xrightarrow{}$ they and `re).

%%%%%%%%%%%%%%%%%%%%%%%%%%%%%%%%%%%%%%%%%%%%%%%%%%%%%%%%%%%%%%%%%%%%%%%%%%%%
% Attention behavior
%%%%%%%%%%%%%%%%%%%%%%%%%%%%%%%%%%%%%%%%%%%%%%%%%%%%%%%%%%%%%%%%%%%%%%%%%%%%
\subsection{Attention Behavior}

%%%%%%%%%%%%%%%%%%%%%%%%%%%%%%%%%%%%%%%%%%%%%%%%%%%%%%%%%%%%%%%%%%%%%%%%%%%%
% SWBD line attention Diagram 
%%%%%%%%%%%%%%%%%%%%%%%%%%%%%%%%%%%%%%%%%%%%%%%%%%%%%%%%%%%%%%%%%%%%%%%%%%%%
\begin{figure}
\centering
\includegraphics[width=\linewidth]{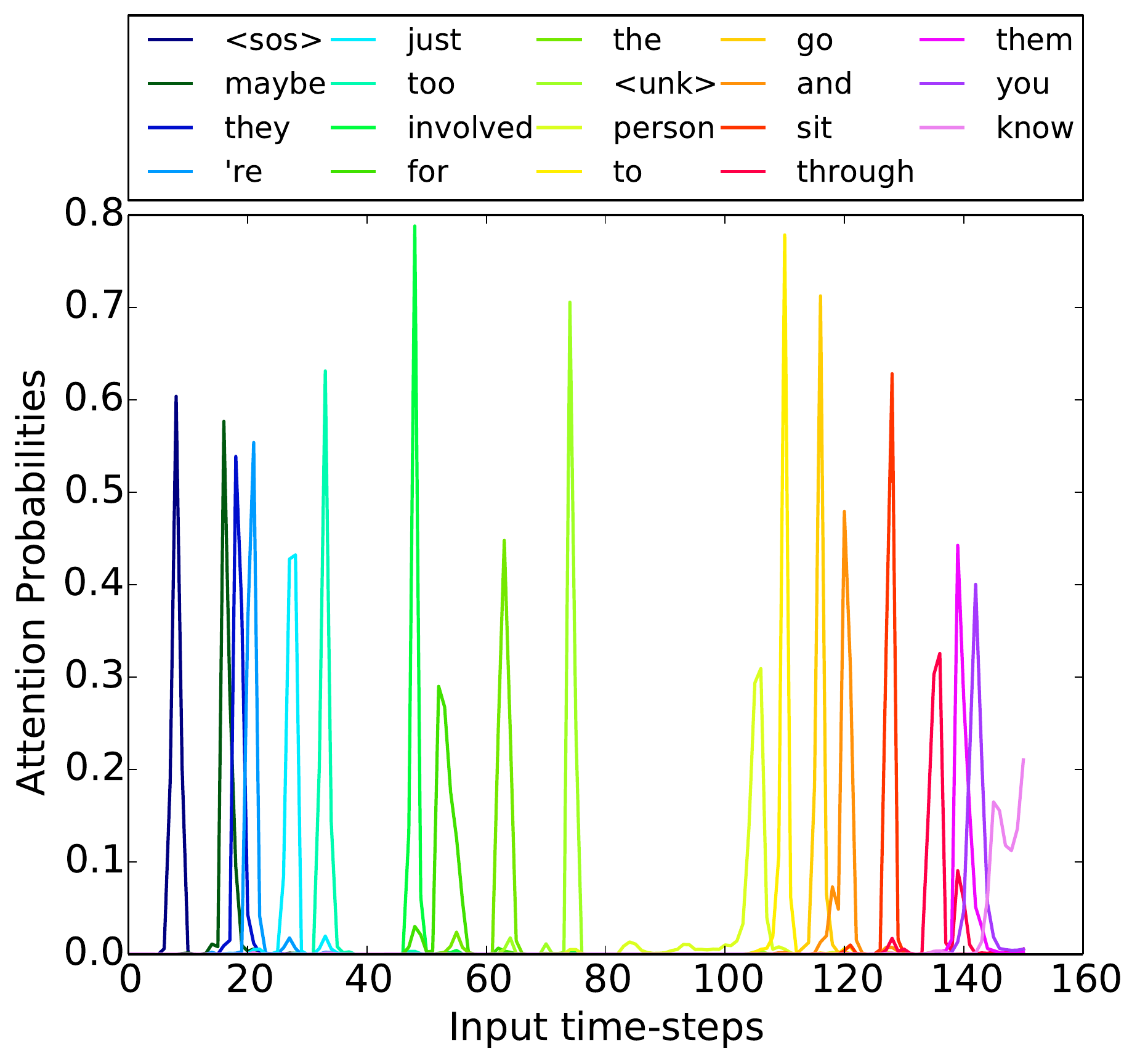}
\caption{Attention visualization for a sample utterance from the validation set shows highly localized attention for a word-level S2S model}
\label{fig:swbd_attn_line_plot}
\end{figure}
%%%%%%%%%%%%%%%%%%%%%%%%%%%%%%%%%%%%%%%%%%%%%%%%%%%%%%%%%%%%%%%%%%%%%%%%%%%%

In Figure \ref{fig:swbd_attn_line_plot} we plot the attention of a sample utterance from our validation set of the corpus. We notice that the attention is very peaky and focuses only on certain frames in the input although generally a word spans multiple input frames. 

To understand this behavior of the model, let us revisit the location-aware attention explained in Section \ref{sec:methods}. The location-aware attention is useful in speech to enforce a monotonic alignment between source and target. It does so by convolving the previous attention vector along input time-steps and feeding it as another input parameter while calculating attention of the current time step. This way, the model is informed where to pay attention ``next'' and would mostly look in the ``future'' to make a prediction. %We see that the attention probability is much lower when the model produces an $<$\textit{unk}$>$ token but the model recovers quickly.

As this model is trained towards word-units and the attention is focused only on certain frames, we speculate that the hidden states corresponding to those frames are the speech-word-vectors for those words. Here, we are able to extract speech-word-vectors from an end-to-end model trained for direct word recognition without the need of any predefined forced-alignments. The size of these embeddings is equal to the number of RNN cells in the last layer of the encoder. %We note that there is an area of research where unsupervised methods are used to extract such acoustic-word-embeddings \cite{kamper_acoustic_word_embeddings, speech2vec} but they try to do so without speech recognition. Our work is not comparable with theirs. Here we present a method to extract similar embeddings using supervised tasks.

%%%%%%%%%%%%%%%%%%%%%%%%%%%%%%%%%%%%%%%%%%%%%%%%%%%%%%%%%%%%%%%%%%%%%%%%%%%%
% Model learns to segment speech
%%%%%%%%%%%%%%%%%%%%%%%%%%%%%%%%%%%%%%%%%%%%%%%%%%%%%%%%%%%%%%%%%%%%%%%%%%%%
\subsection{Automatic Segmentation of Speech into Words}

Given that the attention is highly localized, we attempt to quantify whether the attention weights corresponded to actual word boundaries. From the Switchboard NXT dataset, we chose all utterances (train, validation and test) for which we have 0\% WER during testing. 39\% of the total utterances have 0 WER. We perform decoding with beam size 1 here. We converted the human-annotated forced-alignments to their corresponding frame numbers using the 10ms frame rate of our model. The predicted frame number is calculated from the attention distribution shown in Figure \ref{fig:swbd_attn_line_plot} as follows. The input frame with the max attention probability is chosen as the predicted frame for the word. The frame error is calculated at each word level by taking an absolute difference between the predicted and grouthtruth frame number. A positive difference means the predicted frame was after the groundtruth alignment, and a negative difference means that it was before. We average this frame error for all words in all utterances (171073 words). An example of this computation is $\textnormal{Predicted} = [988,1008,1012,1044,1092] ~\\ \textnormal{Groundtruth} = [988,1005,1013,1042,1100] ~\\ \textnormal{and Frame Error} = [0,+3,-1,+2,-8]$. % shown in Table \ref{tab:frame_error_example}.

%%%%%%%%%%%%%%%%%%%%%%%%%%%%%%%%%%%%%%%%%%%%%%%%%%%%%%%%%%%%%%%%%%%%%%%%%%%%
% Frame Errors Example
% 0, 988, 988, 0, well, well 
% 2, 1008, 1005, 3, i'm, 'm  
% 3, 1012, 1013, -1, a, a
% 4, 1044, 1042, 2, school, school
% 5, 1092, 1100, -8, teacher, teacher
%%%%%%%%%%%%%%%%%%%%%%%%%%%%%%%%%%%%%%%%%%%%%%%%%%%%%%%%%%%%%%%%%%%%%%%%%%%%
% \begin{table}[h!]
% \caption{Frame Error (Err) calculation between predicted frame (Pred) and groundtruth (GT) for each decoder step (\#)} 
% \centering
% \begin{tabular}{ p{1cm}  p{1cm} p{1cm} p{1cm} }
% \hline
% \hline
% \#          & Pred          & GT            &  Err          \\ 
% \hline
% \hline
% 1           & 988           & 988           & 0             \\
% 2           & 1008          & 1005          & +3            \\
% 3           & 1012          & 1013          & -1            \\
% 4           & 1044          & 1042          & +2            \\
% 5           & 1092          & 1100          & -8            \\
% \hline
% \hline
% \end{tabular}
% \label{tab:frame_error_example}
% \end{table}
%%%%%%%%%%%%%%%%%%%%%%%%%%%%%%%%%%%%%%%%%%%%%%%%%%%%%%%%%%%%%%%%%%%%%%%%%%%%

The attention weights for the last word predicted in the sequence is often most erroneous. As an example, in Figure \ref{fig:swbd_attn_line_plot} we see that ``know'', the last word, has a distributed attention weight, and has the least probability value (approximately 0.2) compared to other words. For better understanding, we also compute frame errors without considering the last word of every utterance.

We compute the mean and standard deviation of frame errors for all words. During training, we use a pyramidal encoder that reduces the input frame lengths by a factor of 4. Hence, while computing mean and standard deviation of frame errors, we scale them by 4 as well for fair comparison. The standard deviation of frame error without including last word is 3.6 frames after the groundtruth. For a word-based model, this is an encouraging result as usually a character unit spans 7 (or 1.75 frames after a pyramidal encoder) and a word would span many more.

%%%%%%%%%%%%%%%%%%%%%%%%%%%%%%%%%%%%%%%%%%%%%%%%%%%%%%%%%%%%%%%%%%%%%%%%%%%%
% Frame Errors
% showing scaled errors by 4
%%%%%%%%%%%%%%%%%%%%%%%%%%%%%%%%%%%%%%%%%%%%%%%%%%%%%%%%%%%%%%%%%%%%%%%%%%%%
\begin{table}[h!]
\caption{Average frame error mean and standard deviation (std dev.) between groundtruth forced-alignments and S2S word segment prediction} 
\centering
\begin{tabular}{ p{3.5cm}   p{0.8cm}  p{0.8cm} p{0.8cm}}
\hline
\hline
\hspace*{0.2cm}
			    & \multicolumn{3}{c}{Avg. Frame Error} \\
			    &  Train 	& 	Val 		& Test \\ \hline
            \hline
W/o Last Word - Mean     & 0	  &  -0.08	  & -0.01\\
W/o Last Word - Std Dev & 3.7  &   3.3     & 2.0 \\
All Words - Mean     & 0.4  & 	0.3	& 0.3	\\
All Words - Std Dev & 10.1 &  9.8 & 10.5 \\
\hline
\hline
\end{tabular}
\label{tab:frame_err_computation}
\end{table}
%%%%%%%%%%%%%%%%%%%%%%%%%%%%%%%%%%%%%%%%%%%%%%%%%%%%%%%%%%%%%%%%%%%%%%%%%%%%

%%%%%%%%%%%%%%%%%%%%%%%%%%%%%%%%%%%%%%%%%%%%%%%%%%%%%%%%%%%%%%%%%%%%%%%%%%%%
% add why attention focuses on end of word
%%%%%%%%%%%%%%%%%%%%%%%%%%%%%%%%%%%%%%%%%%%%%%%%%%%%%%%%%%%%%%%%%%%%%%%%%%%%
\textbf{Why does attention focus on the end of word?} The optimization task in A2W recognition is to map a sequence of input frames (usually larger number of input frames than in character or BPE prediction models) to a sequence of target words. During training, the model learns where word boundaries occur by recognizing the attention distribution that leads to highest probability of generating the correct output. The bi-directional LSTM in the encoder has access to the past as well as future input. Therefore, the encoder learns to look into the future to recognize where a different word is beginning, and the BLSTM would hold richest embeddings in the unit corresponding to each of frame of the current word. We investigate the encoder embeddings in the next section in more detail. It is also important to note that the location-aware attention constrains the model to only look into the future, and not the past, which would push the boundaries towards word ends rather than beginnings. Hence, the attention mechanism learns to focus mostly on the word boundaries.

We obtain a context vector from the attention mechanism that is a weighted sum of the encoder hidden states. Following this peaky nature of the attention mechanism, we expect to see certain patterns reflected in the encoder embeddings. This is explored in the following section.

\section{Speech Embeddings}

%%%%%%%%%%%%%%%%%%%%%%%%%%%%%%%%%%%%%%%%%%%%%%%%%%%%%%%%%%%%%%%%%%%%%%%%%%%%%
% fixed utterance sw3583BT (from the parse train_dev set)
% say that people first needed to use complicated methods or run rrns over speech to extract neural embeddings
%%%%%%%%%%%%%%%%%%%%%%%%%%%%%%%%%%%%%%%%%%%%%%%%%%%%%%%%%%%%%%%%%%%%%%%%%%%%

%%%%%%%%%%%%%%%%%%%%%%%%%%%%%%%%%%%%%%%%%%%%%%%%%%%%%%%%%%%%%%%%%%%%%%%%%%%%
% Speech Embed diagram 
%%%%%%%%%%%%%%%%%%%%%%%%%%%%%%%%%%%%%%%%%%%%%%%%%%%%%%%%%%%%%%%%%%%%%%%%%%%%
\begin{figure*}[ht!]
\centering
    \begin{minipage}{\textwidth}
    \begin{center}
    \includegraphics[width=1.0\linewidth, height=12cm]{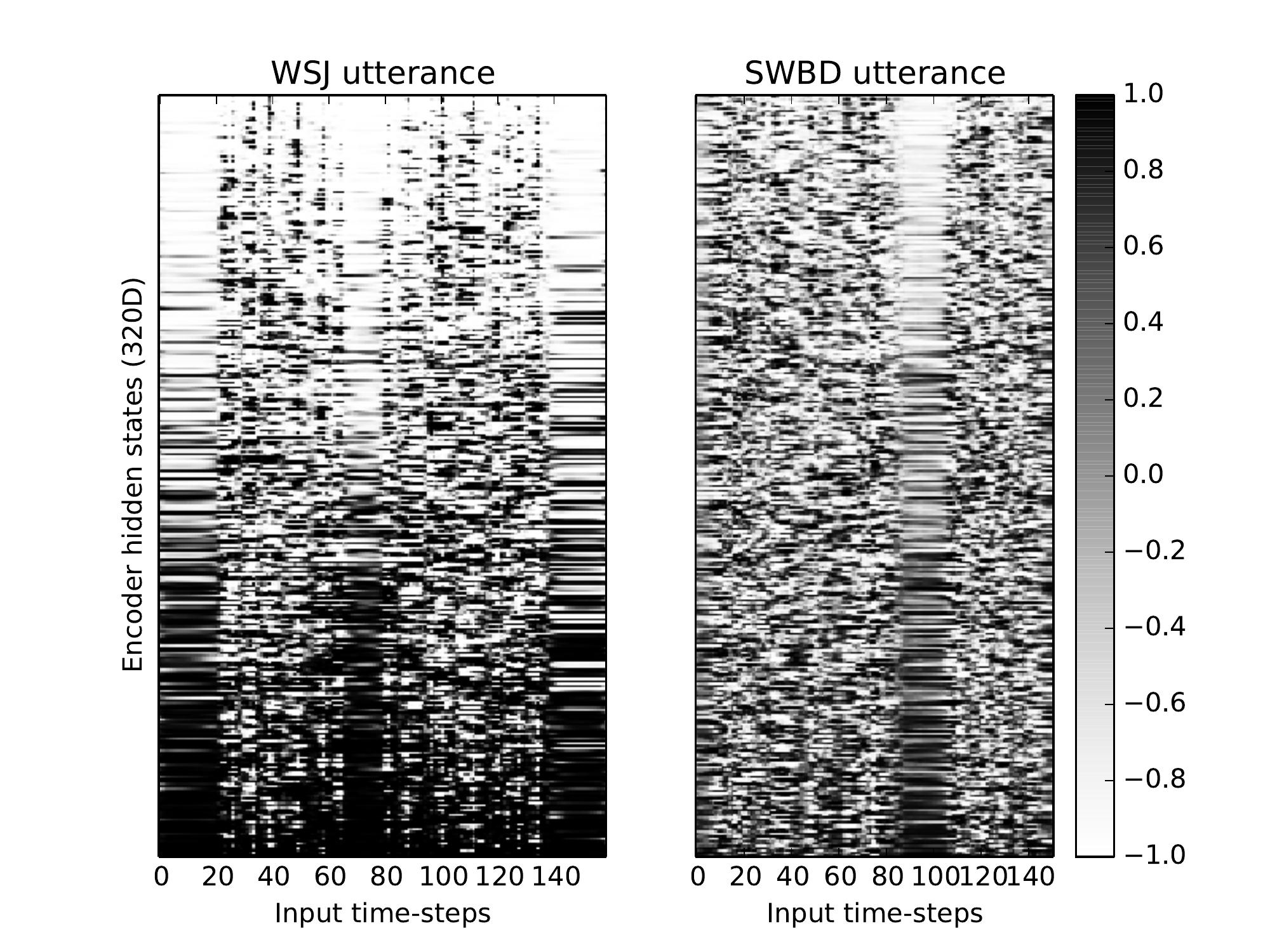} \vspace{-1.7em}
    \end{center}
    \end{minipage}
    {\hfill}
    \vspace{-1.4em}
    
\vspace{1em}
\caption{Encoder hidden state visualization for WSJ (acoustically clean data) and SWBD (acoustically noisier). Visualization shows encoder activations across input time frames.}
\label{fig:hidden_vector_visualization}
\end{figure*}
%%%%%%%%%%%%%%%%%%%%%%%%%%%%%%%%%%%%%%%%%%%%%%%%%%%%%%%%%%%%%%%%%%%%%%%%%%%%
We train a similar A2W model on the Wall Street Journal corpus (WSJ, LDC93S6B and LDC94S13B) which comprises about 90 hours of read speech in clean acoustic environments with a close-talk microphone. This dataset has about 300 different speakers in the train, validation (dev93) and test (eval92) sets. WSJ is sampled at 16kHz while SWBD is sampled at 8kHz and we upsample SWBD to 16kHz for implementation reasons. We bring the readers attention to these major differences in acoustic and speaker variability and domain of the data in WSJ and SWBD. In Figure \ref{fig:hidden_vector_visualization} we visualize the encoder hidden states for sample utterances from the validation sets of WSJ (4k8c030h) and SWBD (same as in Figure \ref{fig:swbd_attn_line_plot}). We train a WSJ model to compare hidden state activations of the noisier SWBD dataset with a clean WSJ dataset as we expect the activation patterns to be clearer and more interpretable in the cleaner dataset. The hidden state dimension here is same as the number of BLSTM cells in the last encoder layer (320D). For this visualization, we sort the hidden states of the encoder in an ascending order of total activation over time. We use a tanh non-linearity hence all values range from -1 to +1. We note that there are three types of patterns to observe in these activations: 1) stable horizontal lines, 2) disruptions, and 3) vertical dashed-line pattern across encoder hidden states (Y-axis) within the disruptions. Pattern 3 is easier to notice in the WSJ activations.

Upon listening to these utterances, we found that stable horizontal lines (pattern 1) corresponds to silence in the utterance, while disruptions (pattern 2) corresponds to the speech. We observe similar patterns identifying speech and non-speech in both WSJ and SWBD. From this, we understand that the model has learned to detect and segment pauses in speech. As WSJ is the acoustically cleaner corpus with less variability, ``silence'' acoustics are stable and repetitive throughout, which is what we observe in the beginning, middle and end of the WSJ utterance--while the SWBD ``silence'' activations look different. In WSJ, we can further identify multiple vertical dashed-line patterns across all encoder hidden states (i.e. Y-axis; pattern 3). This pattern is formed by encoder units turning on and off (+1, -1) when a word boundary is reached. This particular WSJ utterance has 15 words and we observe 15 vertical dashed-line patterns in the activations. This further reinstates that we are able to represent multiple frames of speech using single 320D speech-word-vectors. Pattern 3 is tougher to spot in SWBD comparatively but still noticeable; it might need more training data or better regularization with this data to obtain similar properties as the WSJ model.

% These input time-steps may correspond to those inputs in the attention visualization \ref{fig:swbd_attn_line_plot} where attention is peaky.
%%%%%%%%%%%%%%%%%%%%%%%%%%%%%%%%%%%%%%%%%%%%%%%%%%%%%%%%%%%%%%%%%%%%%%%%%%%

\section{Conclusion}

In this paper, we presented promising results on character-based and word-based S2S models on the 300 hour Switchboard corpus with improved training strategies. We then show a quantitative analysis of model behavior by analyzing the encoder hidden states and attention mechanism. We find that the model learns to segment speech into word, silence, and non-silence parts without any supervision other than word-level transcripts (with utterance level alignments). We also show that it is possible to extract speech-word-vectors from this type of model. As a follow up study, we would like to explore this behavior on corpora other than Switchboard or Wall Street Journal.

\section{Acknowledgements}
\label{sec:acknowledgements}
We thank Desmond Elliot, Amanda Duarte, Ozan Caglayan and Jindrich Libovicky for their valuable feedback on this writeup. We also thank the CMU speech group for many useful discussions. We gratefully acknowledge the support of NVIDIA Corporation with the donation of the Titan X Pascal GPUs used for this research. This work is partially supported by the DARPA AIDA grant.
% Todo: acknowledge maybe Amazon, Google, or Facebook - need to check. Maybe DARPA AIDA? Pittbsurgh Supercomputing Center?

\section*{Appendix}
\begin{figure*}[ht!]
\centering
    \begin{minipage}{\textwidth}
    \begin{center}
    \includegraphics[width=1.0\linewidth]{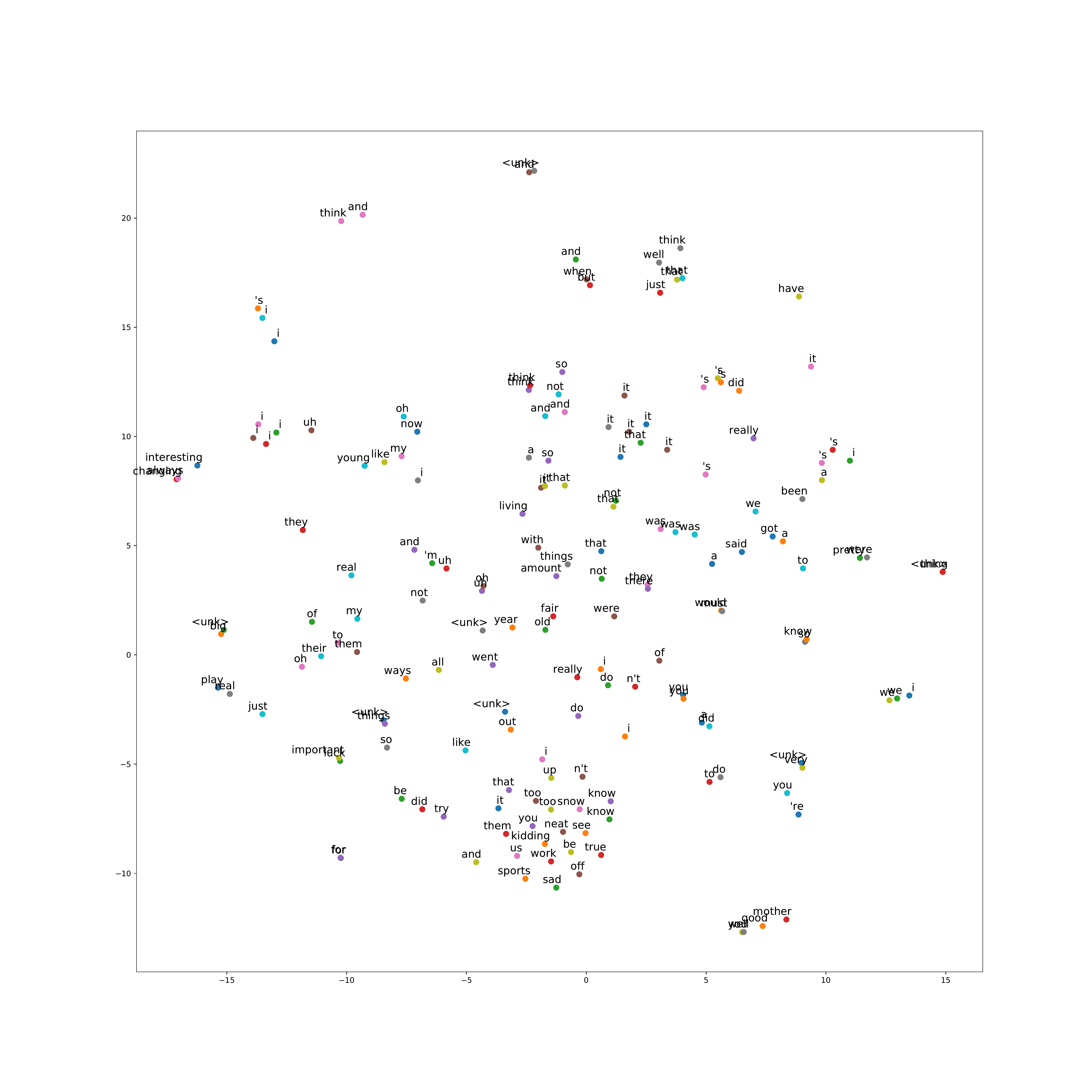} \vspace{-1.7em}
    \end{center}
    \end{minipage}
    {\hfill}
    \vspace{-1.4em}
    
\vspace{1em}
\caption{TSNE visualization of 300 randomly selected speech embeddings. We observe that same words occur close together which shows that the embeddings are good representations of the words.}
\label{fig:TSNE_visualization}
\end{figure*}

In this section, we investigate the Speech Embeddings further using TSNE visualization and finding nearest neighbors of each acoustic-word-embedding. In Figure \ref{fig:TSNE_visualization} and Table \ref{tab:Nearest_neighbors} we see ``same'' words cluster together.

%%%%%%%%%%%%%%%%%%%%%%%%%%%%%%%%%%%%%%%%%%%%%%%%%%%%%%%%%%%%%%%%%%%%%%%%%%%%
% TSNE visualization 
%%%%%%%%%%%%%%%%%%%%%%%%%%%%%%%%%%%%%%%%%%%%%%%%%%%%%%%%%%%%%%%%%%%%%%%%%%%%
%%%%%%%%%%%%%%%%%%%%%%%%%%%%%%%%%%%%%%%%%%%%%%%%%%%%%%%%%%%%%%%%%%%%%%%%%%

%%%%%%%%%%%%%%%%%%%%%%%%%%%%%%%%%%%%%%%%%%%%%%%%%%%%%%%%%%%%%%%%%%%%%%%%%%%%
% Nearest neighbors
% 
%%%%%%%%%%%%%%%%%%%%%%%%%%%%%%%%%%%%%%%%%%%%%%%%%%%%%%%%%%%%%%%%%%%%%%%%%%%%
\begin{table*}[h!]
\caption{Nearest Neighbor search over acoustic-word-embeddings. Table shows 10 nearest neighbor for a particular word. Words shown below are randomly chosen.} 
\centering
\begin{tabular}{ p{1cm} p{10cm}   }
\hline
\hline
word & nearest neighbors \\
\hline
oh & oh, oh, oh, oh, oh, oh, oh, oh, \#eos\#, oh \\
i & i, i, i, i, i, i, i, i, i, i\\
see & see, see, see, see, see, see, see, see, see, see \\
\#eos\# & \#eos\#, \#eos\#, \#eos\#, \#eos\#, \#eos\#, \#eos\#, \#eos\#, \#eos\#, \#eos\#, \#eos\# \\
that & that, that, that, that, that, that, neat, obviously, that, it \\
's & 's, 's, \#eos\#, 's, 's, 's, you, 's, 's, 's \\
really & really, know, a, so, really, really, really, really, well, really \\
neat & neat, me, neat, \#eos\#, neat, neat, neat, neat, \#eos\#, funny \\
\#eos\# & \#eos\#, \#eos\#, \#eos\#, \#eos\#, \#eos\#, well, \#eos\#, she, \#eos\#, \#eos\# \\
and & and, and, i, and, and, \#eos\#, and, and, and, and \\
so & so, so, so, so, so, couple, know, so, well, so \\
we & we, they, we, we, \#eos\#, we, we, go, we, they \\
did & did, just, just, tell, just, just, in, just, because, goes \\
you & you, you, east, of, you, you, you, you, thing, you \\
know & know, know, know, know, know, know, know, know, would, know \\
\hline
\hline
\end{tabular}
\label{tab:Nearest_neighbors}
\end{table*}
%%%%%%%%%%%%%%%%%%%%%%%%%%%%%%%%%%%%%%%%%%%%%%%%%%%%%%%%%%%%%%%%%%%%%%%%%%%%

\bibliographystyle{IEEEbib}
\bibliography{refs}

\end{document}